\documentclass[12pt]{article}
\usepackage{axodraw}
\usepackage{amssymb}

\begin{document}
\begin{titlepage}
\begin{center}
\hfill    CERN-PH-TH/2008-183\\
\hfill    NORDITA-2008-38 \\

\vskip 1cm


{\large \bf {Building the full PMNS Matrix from six independent
Majorana-type phases}}

\vskip 1cm

Gustavo C. Branco$^{a, b}$,\footnote{On sabbatical leave at 
Universitat de Val\` encia-CSIC until 30 June 2008. \\
Email: gbranco@ist.utl.pt} 
M. N. Rebelo$^{a, c, d}$  \footnote{On sabbatical leave at CERN
PH-TH during part of 2007/2008. Core scientist at
the NORDITA program ``TeV scale physics and dark matter" 
Summer 2008. \\
Email: margarida.rebelo@cern.ch  and rebelo@ist.utl.pt}
\vskip 0.05in

{\em a Departamento de F{\'\i}sica and Centro  de F{\'\i}sica
Te{\'o}rica de Part{\'\i}culas (CFTP),\\
Instituto Superior T\'{e}cnico (IST), Av. Rovisco Pais, 1049-001
Lisboa, Portugal. \\
b Departament de F\' \i sica Te\` orica and IFIC,
Universitat de Val\` encia-CSIC, E-46100, Burjassot, Spain.\\
c CERN, Department of Physics, Theory Unit, CH-1211, Geneva 23, 
Switzerland. \\
d NORDITA, Roslagstullsbacken 23, SE-10691, Stockholm, Sweden.
}

\end{center}

\vskip 3cm

\begin{abstract}
In the framework of three light Majorana neutrinos, we show how to 
reconstruct, through the use of $3 \times 3$ unitarity, 
the full PMNS matrix from 
six independent Majorana-type phases. In particular, we express
the strength of Dirac-type CP violation in terms of these
Majorana-type phases by writing the area of the unitarity
triangles in terms of these phases.
We also study how these six
Majorana phases appear in CP-odd weak basis invariants as well as in leptonic 
asymmetries relevant for flavoured leptogenesis.

\end{abstract}

\end{titlepage}

\newpage
\section{Introduction}
The discovery of neutrino oscillations \cite{kayser} 
providing evidence for 
non-vanishing neutrino masses and leptonic mixing, is one of 
the most exciting recent developments in Particle Physics. 
At present, it is not known whether neutrinos
are Dirac or Majorana fermions. The latter possibility has the special appeal 
of providing, through the seesaw mechanism \cite{seesaw1}--- \cite{seesaw5}
an elegant explanation of why 
neutrinos are much lighter than the other known fermions. It is well known
that the presence of Majorana neutrinos introduces some novel features
in leptonic CP violation, like the possibility of having CP violation  
in the case of two Majorana neutrinos as well as having CP breaking even
in the limit of three exactly degenerate neutrinos \cite{Branco:1998bw}.
These features reflect the fact that in the presence of Majorana neutrinos, 
the simplest non-trivial rephasing invariant functions 
of the leptonic mixing matrix elements, are bilinears and not quartets, 
as it is the case for Dirac particles. 
We designate ``Majorana-type phases" the arguments of these 
rephasing invariant bilinears. Physically, these phases correspond 
to the orientation in the complex plane of the sides of 
the Majorana unitarity triangles. Recall that
in the case of the quark sector and in general for Dirac particles,
the orientations of the unitarity triangles have no physical meaning,
reflecting the fact that Dirac unitarity triangles rotate under rephasing
of the quark fields.
The leptonic mixing matrix, often called Pontecorvo, Maki, Nakagawa and 
Sakata (PMNS) matrix is usually parameterised by a $3 \times 3$ unitary 
matrix containing three
mixing angles, one Dirac phase and two Majorana phases, for a total of six
independent parameters. It should be emphasized that these Majorana phases 
are related to but do not coincide with the above defined 
Majorana-type phases. The crucial point is that Majorana-type 
phases are rephasing invariants which are measurable quantities,
and do not depend on any particular parameterisation of the PMNS matrix.

In this paper we adopt the arguments of these rephasing invariant bilinears 
as fundamental parameters and we show that, in the framework of 
three light Majorana neutrinos and in presence of Dirac-type CP
violation, the full PMNS matrix can be reconstructed from six independent 
Majorana-type phases. We also study how these Majorana-type phases appear 
in neutrinoless double beta decay, as well as in CP-odd weak basis 
invariants and in 
leptonic asymmetries relevant for flavoured leptogenesis. We conclude
that, in this framework, all low energy leptonic physics is
encoded into six leptonic masses and six Majorana-type phases. 
In the case of one massless neutrino one of these Majorana-type 
phases may be fixed (e.g., chosen to be equal to zero), without
changing the lengths of the sides and internal angles of the 
unitary triangles.\\
 
This paper is organised as follows. In section 2 we set 
the notation and present our framework. 
In section 3, we choose six independent Majorana-type phases 
and show how the full unitary PMNS matrix can be constructed 
from these six input phases. We also show how to express the 
strength of Dirac-type CP violation in terms of the six 
Majorana-type phases and analyse the unitarity triangles,
taking into account the present experimental data.
In sections 4, 5, 6 we show how Majorana-type phases appear
in the elements of the effective mass matrix, in CP-odd 
weak-basis invariants and in leptogenesis, respectively. Finally
our conclusions are contained in section 7.

\section{Framework and Notation}
We consider an extension of the Stardard Model (SM) consisting of the 
addition of an arbitrary number of righthanded neutrinos  leading to 
three light Majorana neutrinos, through the seesaw mechanism. The leptonic 
mixing matrix $V$ is a $3 \times (3+n_R)$ matrix connecting the charged 
leptons to the three light neutrinos and the $n_R$ heavy neutrinos. This
mixing matrix $V$ is, of course, a submatrix
of a $(3+n_R) \times (3+n_R)$ unitary matrix. In this work, we 
are specially interested in the low energy limit of the theory,
where the leptonic mixing matrix reduces to the $3 \times 3$ 
PMNS matrix connecting charged leptons to the light neutrinos. Let 
us choose, for the low energy limit, the physical basis where both the 
charged lepton mass matrix, $m_l$  and the  neutrino mass 
matrix $m_\nu$  are diagonal and real:
\begin{eqnarray}
m_l  = {\rm diag}\  (m_e, m_\mu, m_\tau),  \nonumber \\
m_\nu ={\rm diag} \ (m_1, m_2, m_3)
\label{eq1}   
\end{eqnarray}
In this  basis, there is still the freedom to rephase the charged lepton 
fields:
\begin{equation}
l_j \rightarrow l^\prime_j = {\rm exp}\  (i \phi_j) \ l_j
\label{eq2}
\end{equation}
with  arbitrary $\phi_j$s, which leaves the charged lepton mass terms 
$m_j \overline{l_j} l_j$ invariant. Due to the Majorana nature of the
neutrinos the rephasing:
\begin{equation}
\nu_k \rightarrow \nu^\prime_k = {\rm exp}\  (-i \psi_k ) \ \nu_k
\label{eq3}
\end{equation}
with  arbitrary $\psi_k$s is not allowed, since it would not keep the Majorana 
mass terms $ \nu_{Lk}^{T} C^{-1} m_k \nu_{Lk}$ invariant. Note however 
that one can still make the rephasing of Eq.~(\ref{eq2}) for
$\psi_k = (n_k \pi )$ with $n_k$ an integer.

In the mass eigenstate basis, the  low energy
weak charged current can be written as:
\begin{equation}
{\cal L}_W = - \frac{g}{\sqrt{2}}  \overline{l_{jL}}
\gamma_{\mu} U_{jk} {\nu_k}_L + h.c.,
\label{lw}
\end{equation}
where
\begin{equation}
U = \left[
\begin{array}{ccc}
 U_{e1} & U_{e2} & U_{e3} \\ 
 U_{\mu 1} & U_{\mu 2} & U_{\mu 3} \\ 
 U_{\tau 1} & U_{\tau 2} & U_{\tau 3} \end{array} \right].
\label{uli}
\end{equation}
So far, we have not introduced the constraints of unitarity. 
As a result, $U$ is characterized
by nine moduly and six phases, since three of the nine phases of
$U$ can be eliminated through the rephasing of Eq.~(\ref{eq2}).
If we assume $3 \times 3$ unitarity, it is well known that 
$U$ is characterized
by six parameters which, as mentioned above, are usually taken as
three mixing angles and three CP violating phases.

\section{Reconstruction of the full unitary PMNS matrix from six  
Majorana-type phases}
The study of rephasing invariant quantities is of special importance
for the analysis of mixing and CP violation both in the quark 
and lepton sectors. In the quark sector, the simplest rephasing
invariant quantities are the nine moduli of the elements of the 
Cabibbo-Kobayashi-Maskawa matrix, $V_{CKM}$, and the arguments
of rephasing invariant quartets, like for example, 
$\arg \ (V_{us} V_{cb}V^*_{ub}V^*_{cs})$. The assumption of $3 \times 3$
unitarity of  $V_{CKM}$ leads to a series of exact relations
among various rephasing invariant quantities \cite{Botella:2002fr}
which provide an important test of the SM. Unitarity also allows
for various parameterisations of  $V_{CKM}$ which can be taken as
three moduli and one invariant phase, as in the so-called
standard parameterisation \cite{Amsler:2008zz}, four independent moduli
\cite{Branco:1987mj}, or four independent invariant phases
\cite{Aleksan:1994if}. The novel feature of the low energy limit of
the leptonic sector
with Majorana neutrinos is the existence of rephasing invariant 
bilinears of the type $U_{l\alpha } U^*_{l\beta }$ where 
$ \alpha \neq \beta$ and no summation on repeated indices is implied.
We designate $\arg \ (U_{l\alpha } U^*_{l\beta })$ ``Majorana-type phases''.
These are the minimal CP violating quantities in the case of Majorana
neutrinos \cite{Branco:1986gr} --- \cite{Jenkins:2007ip}. 
Note that in order for these phases to be 
precisely defined we work with real, nonzero neutrino masses
corresponding to Majorana fields which satisfy
Majorana conditions that do not contain phase factors. It can be
readily seen, from their definition, 
that there are only six independent Majorana-type phases even 
in the general case where unitarity is not imposed on $U$. 
All the other Majorana-type phases in $U$ can be obtained 
from these six phases. This reflects 
the freedom one has to rephase the three charged lepton
fields. This would still be true for the matrix $U$ in a general 
framework including an arbitrary number of right-handed neutrinos
($n_R \geq 3$) together with, for instance, an arbitrary number of 
vector-like charged leptons.

We choose the six independent Majorana-type phases to be:
\begin{eqnarray}
{\beta _1} \equiv  \arg \ (U_{e1} U^*_{e2}), \qquad
{\beta _2} \equiv  \arg \ (U_{\mu 1} U^*_{\mu 2}), \qquad
{\beta _3} \equiv  \arg \ (U_{\tau 1} U^*_{\tau 2}), \nonumber
\\
{\gamma _1} \equiv  \arg \ (U_{e1} U^*_{e3}), \qquad
{\gamma _2} \equiv  \arg \ (U_{\mu 1} U^*_{\mu 3}), \qquad
{\gamma _3} \equiv  \arg \ (U_{\tau 1} U^*_{\tau 3}).
\end{eqnarray} 

Let us now consider Dirac-type phases, which correspond to 
the arguments of rephasing invariant
quartets. It can be readily seen that the $3 \times 3$ $U$ matrix
contains four independent Dirac-type phases. Again, this result is 
completely general, in particular it does not depend on the number of 
right-handed neutrinos ($n_R \geq 3$) or the eventual presence of 
vector-like charged leptons. We choose the following four independent
Dirac-type invariant phases:
\begin{eqnarray}
{\sigma }^{12}_{e\mu } \equiv \arg \ (U_{e1} U_{\mu 2}U^*_{e2}
U^*_{\mu 1})  \label{arr1} \\
{\sigma }^{12}_{e\tau } \equiv \arg \ (U_{e1} U_{\tau 2}U^*_{e2}
U^*_{\tau 1})   \\
{\sigma }^{13}_{e\mu } \equiv \arg \ (U_{e1} U_{\mu 3}U^*_{e3}
U^*_{\mu 1})  \\
{\sigma }^{13}_{e\tau } \equiv \arg \ (U_{e1} U_{\tau 3}U^*_{e3}
U^*_{\tau 1})   \label{arr4} 
\end{eqnarray} 
It is clear that these four Dirac-type phases can be obtained from the
six Majorana-type phases:
\begin{eqnarray}
{\sigma }^{12}_{e\mu } = {\beta _1} - {\beta _2} \\
{\sigma }^{12}_{e\tau } = {\beta _1} - {\beta _3} \\
{\sigma }^{13}_{e\mu } = {\gamma _1} - {\gamma _2} \\
{\sigma }^{13}_{e\tau } = {\gamma _1} - {\gamma _3}
\end{eqnarray} 

It follows from these expressions that, in the framework of
Majorana neutrinos, Dirac-type CP violation in the leptonic sector,
necessarily implies Majorana-type CP violation. \\

Now, we assume unitarity of the $3 \times 3$
PMNS matrix and show that in this limit,
it is possible to fully reconstruct the unitarity mixing
matrix from the six Majorana-type phases, ${\beta _j}$, ${\gamma _j}$
provided there is Dirac-type CP violation. This can be shown
making use of the standard parameterisation \cite{Amsler:2008zz}:
\begin{equation}
U = \left( 
\begin{array}{ccc}
c_{12}c_{13} & s_{12}c_{13} & s_{13}e^{-i\delta } \\ 
-s_{12}c_{23}-c_{12}s_{23}s_{13}e^{i\delta } & \quad
c_{12}c_{23}-s_{12}s_{23}s_{13}e^{i\delta }\quad & s_{23}c_{13} \\ 
s_{12}s_{23}-c_{12}c_{23}s_{13}e^{i\delta } & 
-c_{12}s_{23}-s_{12}c_{23}s_{13}e^{i\delta } & c_{23}c_{13}
\end{array}
\right) \,\cdot P,  \label{pdg}
\end{equation}
where $c_{ij}\equiv \cos \theta _{ij}\ ,\ s_{ij}\equiv \sin \theta _{ij}\ $,
with all $\theta _{ij}$ in the first quadrant,  $\delta $ 
is a Dirac-type phase
and  $P=\mathrm{diag\ }(1,e^{i\alpha}, e^{i\beta})$ with 
$\alpha $ and $\beta$ denoting the phases associated with the
Majorana character of neutrinos. \\

The extraction of the angles  $\theta _{ij}$ 
and $\delta$ is done through
the unitarity triangles. There are two types of unitarity triangles, 
those obtained by multiplication of two different rows and those
obtained by multiplication of two different columns. It has
been pointed out  \cite{AguilarSaavedra:2000vr} that these triangles 
are fundamentally different. Those of the first type were designated 
as ``Dirac triangles'' and have similar properties to those 
built in the quark sector.
Their orientation has no physical meaning since, under 
rephasing transformations of the charged lepton fields these 
triangles rotate in the complex plane. Those of the second type 
were designated as ``Majorana triangles''. Under the allowed 
rephasing, these triangles do not rotate in the complex 
plane since the orientations of all their sides correspond to 
the arguments of rephasing invariants. As a result, the
orientation of Majorana triangles is physically meaningful
\cite{AguilarSaavedra:2000vr}.
Of course, all the six triangles share a common area  
$A= 1/2 | {\mbox Im} U_{ij} U^*_{kj} U_{kl} U^*_{il}|$
(no sum in repeated indices, $k \neq i$, $l \neq j$).
The Majorana triangles provide the necessary and 
sufficient conditions for CP conservation, to wit, 
vanishing of their common area $A$
and orientation of all collapsed Majorana triangles 
along the direction of the real or imaginary axes.
The vanishing of $A$ implies that the Dirac phase
$\delta$ of the parameterisation of Eq.~(\ref{pdg}) equals zero or $\pi$. 
The three different Majorana triangles are:
\begin{eqnarray}
U_{e1} U^*_{e2} + U_{\mu 1} U^*_{\mu 2} + U_{\tau 1} U^*_{\tau 2} =0 \label{dif1} \\
U_{e1} U^*_{e3} + U_{\mu 1} U^*_{\mu 3} + U_{\tau 1} U^*_{\tau 3} =0 \label{dif2} \\
U_{e2} U^*_{e3} + U_{\mu 2} U^*_{\mu 3} + U_{\tau 2} U^*_{\tau 3} =0 \label{dif3} 
\end{eqnarray}
Some of the general features of the three Majorana triangles
are worth pointing out. The three internal angles of the 
first Majorana triangle corresponding to Eq.~(\ref{dif1}) 
are given by $\pi - (\beta_i - \beta_j)$ with $i \neq j$ both 
indices ranging from 1 to 3.
Similarly, for the internal angles of the second triangle corresponding
to  Eq.~(\ref{dif2}) with $\beta$'s replaced by $\gamma$'s.
From the internal angles of two Majorana triangles 
one can readily obtain the internal angles of the third triangle. 
Obviously, there are only four independent 
combinations of $(\beta_i - \beta_j)$ and $(\gamma_i - \gamma_j)$ 
which can be taken as those given by Eqs.~(\ref{arr1})
to  (\ref{arr4}). The internal angles of the three different 
Dirac triangles  are also given in terms of these four independent
combinations. It is sufficient to know the internal angles of two 
of the triangles in order to know all internal angles 
of all unitarity triangles. 

Next we show how to obtain the full PMNS matrix from the
knowledge of $\beta_i$, $\gamma_i$.
Through the law of sines we obtain:
\begin{eqnarray}
&\tan^2   \theta_{12}&  =  \frac{|U_{e2}|^2}{|U_{e1}|^2}  = \nonumber \\
& = &  \frac{|\sin(\gamma_1 -\gamma_2)| 
|\sin(-\beta_2 + \gamma_2 + \beta_3 -\gamma_3)| 
|\sin(\gamma_1 -\gamma_3)| }
{|\sin(-\beta_1 + \gamma_1 + \beta_2 -\gamma_2)|
 |\sin(\gamma_2 -\gamma_3)| |\sin(-\beta_1 + \gamma_1 + \beta_3 -\gamma_3)|}
\label{int1}
\end{eqnarray}

\begin{eqnarray}
&\tan^2   \theta_{23} & =  \frac{|U_{\mu 3}|^2}{|U_{\tau 3}|^2}  = \nonumber \\
& = &  \frac{|\sin(\gamma_1 -\gamma_3)| 
|\sin(-\beta_1 + \gamma_1 + \beta_3 -\gamma_3)| 
|\sin(\beta_1 -\beta_2)| }
{|\sin(-\beta_1 + \gamma_1 + \beta_2 -\gamma_2)|
 |\sin(\gamma_1 -\gamma_2)| |\sin(\beta_1 - \beta_3 )|}
\label{int2}
\end{eqnarray}

\begin{eqnarray}
&\tan^2   \theta_{13} & \frac{1}{\sin^2 \theta_{12}}
=  \frac{|U_{e3}|^2}{|U_{e2}|^2}  = \nonumber \\
& = &  \frac{|\sin(\gamma_2 -\gamma_3)| 
|\sin(\beta_1 -\beta_3)|
|\sin(\beta_1 -\beta_2)| }
{|\sin(\gamma_1 -\gamma_3)|
 |\sin(\gamma_1 -\gamma_2)| |\sin(\beta_2 - \beta_3 )|}
\label{int3}
\end{eqnarray}

From Eqs.~(\ref{int1}), (\ref{int2}) and   (\ref{int3})
we can easily extract the angles $\theta_{ij}$ from the knowledge
of the Majorana phases. 
Finally the phase $\delta$  can be obtained by computing the common
area of the triangles. For instance, from the second triangle we obtain:
\begin{eqnarray}
A = \frac{1}{2}|U_{e1} U^*_{e3}| |U_{\mu 1} U^*_{\mu 3}|  |\sin(\gamma_1 -\gamma_2)|
\end{eqnarray}
From the law of sines we replace $ |U_{\mu 1} U^*_{\mu 3}|$ by:
\begin{eqnarray}
|U_{\mu 1} U^*_{\mu 3}| =  |U_{e1} U^*_{e3}| 
\frac{|\sin(\gamma_1 -\gamma_3)|}{|\sin(\gamma_2 -\gamma_3)|}
\end{eqnarray}
which leads to:
\begin{eqnarray}
A = \frac{1}{2}|c_{12}\  c_{13}\  s_{13}|^2   |\sin(\gamma_1 -\gamma_2)|
\frac{|\sin(\gamma_1 -\gamma_3)|}{|\sin(\gamma_2 -\gamma_3)|} \label{eq21}
\end{eqnarray}
Since the  $\theta_{ij}$ are obtained from $\beta_i$, $\gamma_i$,
using Eqs.~(\ref{int1}), (\ref{int2}) and   (\ref{int3}), it follows 
that Eq.~(\ref{eq21}) gives us the common area of the triangles,
in terms of Majorana phases.
The phase $\delta$ entering in the
standard parameterisation, is readily obtained by recalling that
$A = 1/2 {\mbox Im} Q$ where $Q$ denotes any rephasing invariant quartet.
One obtains: 
\begin{eqnarray}
A =  \frac{1}{16}| \sin(2\,\theta_{12}) \sin(2\,\theta_{13}) 
\sin(2\,\theta_{23})
\cos(\theta_{13})\sin \delta | \label{eq22}
\end{eqnarray}
From Eqs.~(\ref{eq21}), (\ref{eq22}) one obtains      
$\delta$ in terms of Majorana phases.
The quadrant of $\delta$ and the angles $\alpha$ and $\beta$ of Eq.~(\ref{pdg}) are
obtained by inspection.  \\

\subsection{The Strength of Dirac-type CP Violation}
As we have seen, in the limit of $3 \times 3$ unitarity, the six
Majorana-type phases completely fix the PMNS mixing matrix and 
therefore the strength of Dirac-type CP violation, which is
given by $|\mbox{Im}Q|$ where $Q$ denotes any rephasing invariant quartet 
of the PMNS matrix, like for example $Q=(U_{e2} U_{\mu 3} U^*_{e3}U^*_{\mu 2})$.
Note that in the framework of $3 \times 3$ unitarity, one can infer
the size of $|\mbox{Im}Q|$ even without  the direct measurement 
of any CP violating observable. Indeed, as it was shown for the
quark sector \cite{Branco:1987mj}, $|\mbox{Im}Q|$ can be expressed in terms of
four independent moduli of the PMNS matrix. From the present experimental 
data, one cannot infer the size of Dirac-type leptonic CP violation, 
which can range from zero, for instance in the case of vanishing $U_{e3}$, 
to a significant value, of order $10^{-2}$, therefore 
much larger than the corresponding value in
the quark sector where $|\mbox{Im}Q|({\mbox{quark}}) \sim 10^{-5}$. \\

The explicit expression for $|\mbox{Im}Q|$ in terms of the six 
Majorana-type phases is given by:
\begin{equation}
|\mbox{Im}Q| = I_1 I_2 I_3 I_4 I_5 I_6 I_7 I_8 I_9\ /\  D^2
\end{equation}
with
\begin{eqnarray}
& D =  |\sin(\beta_1 -\beta_2)| |\sin(\beta_1 -\beta_3)|
|\sin(\gamma_2 -\gamma_3)| |\sin(-\beta_2 + \gamma_2 + \beta_3 -\gamma_3)| +  \\ 
&+ |\sin(\gamma_1 -\gamma_2)|  |\sin(\gamma_1 -\gamma_3)| |\sin(\beta_2 -\beta_3)|
 |\sin(-\beta_2 + \gamma_2 + \beta_3 -\gamma_3)| +  \nonumber \\
&+|\sin(-\beta_1 + \gamma_1 + \beta_2 -\gamma_2)|
|\sin(-\beta_1 + \gamma_1 + \beta_3 -\gamma_3)| |\sin(\gamma_2 -\gamma_3)|
 |\sin(\beta_2 -\beta_3)|  \nonumber
\end{eqnarray}
and $ I_1 I_2 I_3 I_4 I_5 I_6 I_7 I_8 I_9$ denoting the product of the
sines of the nine internal angles of the three Majorana triangles,
or else of the three Dirac triangles:
\begin{eqnarray}
& I_1 I_2 I_3 I_4 I_5 I_6 I_7 I_8 I_9 =  |\sin(\beta_1 -\beta_2)|
|\sin(\beta_1 -\beta_3)|  |\sin(\beta_2 -\beta_3)| \times \nonumber \\
& \times |\sin(\gamma_1 -\gamma_2)|  |\sin(\gamma_1 -\gamma_3)|
|\sin(\gamma_2 -\gamma_3)| \times \\
& \times |\sin(-\beta_1 + \gamma_1 + \beta_2 -\gamma_2)|
|\sin(-\beta_1 + \gamma_1 + \beta_3 -\gamma_3)| 
|\sin(-\beta_2 + \gamma_2 + \beta_3 -\gamma_3)| \nonumber
\end{eqnarray}

The case of no Dirac-type CP violation is a singular case, where
all unitarity triangles collapse to a line and the 
matrix $U$ can be written as  a real unitary
matrix with two factored out phases which
are usually called  Majorana phases
in the standard parameterisation.
In this case the phases of  Majorana bilinears decouple 
from the size of mixing angles and Eqs. (\ref{int1}) - (\ref{int3}) 
become indetermination relations of the form $0/0$ due to the equality
modulo $\pi$ among all $\beta_j$ Majorana-type phases as well as equality
modulo $\pi$ of all $\gamma_j$ among themselves. \\

We address now the question of finding the values of the six
fundamental Majorana phases which lead to a maximal value of
$|\mbox{Im}Q|$. It can be readily seen that the following
choice of   ${\beta _j}$, ${\gamma _j}$ leads to a maximal value
of Dirac-type CP violation:
\begin{equation}
\beta _k = \frac{2 \pi}{3} k \: , \qquad \gamma _k =  \frac{4 \pi}{3} k \: ,
\qquad k = 1, 2, 3
\end{equation}
This choice of the six  different Majorana-type phases,
together with the adoption of a specially convenient phase 
convention leads to the following PMNS matrix:
\begin{equation}
U_M = \left(
\begin{array}{ccc}
\frac{1}{\sqrt 3} \: w \quad & \frac{1}{\sqrt 3}\quad & \frac{1}{\sqrt 3}\: w^* \\  
 \frac{1}{\sqrt 3}\: w^*\quad &  \frac{1}{\sqrt 3} \quad
& \frac{1}{\sqrt 3}\:  w \\
\frac{1}{\sqrt 3}  \quad &
\frac{1}{\sqrt 3}  \quad  & \frac{1}{\sqrt 3}
\end{array}
\right)  \label{umax}
\end{equation}
where $w = \exp i\frac{2 \pi}{3}$. All unitarity triangles 
corresponding to $U_M$ are equilateral and the maximal value 
of CP violation corresponds to:
\begin{equation}
|\mbox{Im}Q| = \frac{1}{9} \frac{\sqrt 3}{2}
\end{equation}

\subsection{Unitarity Triangles and Present Experimental Data}
The current experimental bounds on neutrino masses and leptonic
mixing are \cite{Amsler:2008zz}:
\begin{eqnarray}
\Delta m^2_{21} & = & (8.0 \pm 0.3) \times 10^{-5}\  {\rm eV}^2 \\
\sin^2 (2 \theta_{12}) & = & 0.86 ^{+0.03}_{-0.04} \label{sim} \\
|\Delta m^2_{32}| & = & (1.9 \ \  \mbox{to} \ \  3.0) \times 10^{-3}\
{\rm eV}^2 \\
\sin ^2 ( 2 \theta_{23}) & > & 0.92 \\
\sin ^2 (2 \theta_{13}) &  < & 0.19
\end{eqnarray}
with $\Delta m^2_{ij} \equiv m^2_j - m^2_i$, where $m_j$'s denote the 
neutrino masses. The angle $ \theta_{23} $
may be maximal, meaning $45^{\circ}$, whilst
$ \theta_{12} $ is already known to deviate from this value. At the moment
there is an experimental upper bound on the angle $ \theta_{13}$.
Recently, there are hints of $\theta_{13} > 0$ from 
global neutrino data analysis, which provides the global 
estimate \cite{Fogli:2008jx}
\begin{equation}
\sin^2 \theta_{13} = 0.016 \pm 0.010 \quad (1 \sigma)
\end{equation}

Present experimental data suggest that in leading order
the leptonic mixing matrix may be approximated by the
Harrison, Perkins and Scott (HPS) mixing matrix \cite{Harrison:2002er}:
\begin{equation}   
\left[
\begin{array}{ccc}
 \frac{2}{\sqrt 6}  & \frac{1}{\sqrt 3} & 0 \\
 - \frac{1}{\sqrt 6}  &  \frac{1}{\sqrt 3} &  \frac{1}{\sqrt 2} \\
  - \frac{1}{\sqrt 6} & \frac{1}{\sqrt 3} & - \frac{1}{\sqrt 2}
\end{array}
\right]
\label{scott}
\end{equation}
which is often designated as tri-bimaximal mixing and 
corresponds to $ \tan \theta_{12} = 1/ \sqrt{2}$,
 $ \theta_{23} = \pi/4 $ and $ \theta_{13} = 0 $

From the point of view of leptonic low energy phenomenology, 
a value of $\theta_{13}$ not far from
its present experimental bound would have interesting experimental 
implications and would allow for the possibility of Dirac-type CP
violation to be detected experimentally in the near future provided the value 
of the  phase $\delta$ is not suppressed.

We address now the question of what unitarity triangles correspond to a
perturbation of the HPS matrix which consists of keeping the 
values for $\theta_{12}$ and $\theta_{23}$ 
fixed and choosing  $\delta$  and $\theta_{13}$ 
that  maximize the area of the unitarity triangle, with 
 $\theta_{13}$ within  the experimentally  allowed values
(i.e., $\sin \theta_{13} = 0.22$ and $\delta = \pi /2$). It follows from
Eq.~(\ref{pdg}) that this perturbation spoils the exact trimaximal mixing of
the second column of the HPS matrix.
\begin{figure}[t]
\begin{center}
\begin{picture}(400,80)(0,0)
\ArrowLine(20,10)(190,10)
\Text(90,0)[]{$U_{e1} U^*_{e2}$}
\ArrowLine(190,10)(105,51.7)
\Text(170,40)[]{$U_{\tau 1} U^*_{\tau 2}$}
\ArrowLine(105,51.7)(20,10)
\Text(40,40)[]{$U_{\mu 1} U^*_{\mu 2}$}
\ArrowLine(350,10)(350,76.4)
\Text(380,43.2)[]{$U_{e1} U^*_{e3}$}
\ArrowLine(350,76.4)(243.2,43.2)
\Text(260,65)[]{$U_{\mu 1} U^*_{\mu 3}$}
\ArrowLine(243.2,43.2)(350,10)
\Text(260,20)[]{$U_{\tau 1} U^*_{\tau 3}$}
\end{picture}
\end{center}
\caption{First and Second Majorana unitarity triangles, 
corresponding to Eqs.~(\ref{fir}) and (\ref{sec})}
\label{majtri}
\end{figure}
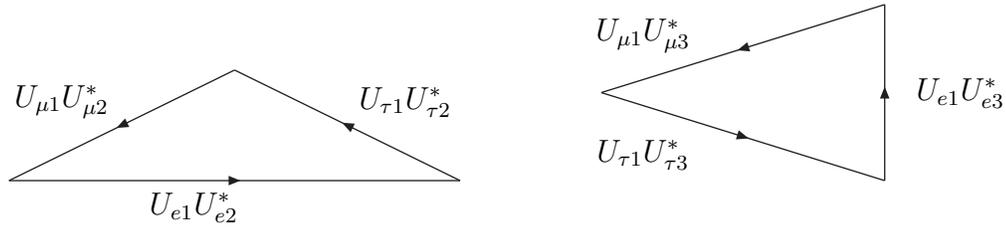
In this case we have for the first Majorana triangle
\begin{equation}
U_{e1} U^*_{e2} = 0.448 \qquad
U_{\mu 1} U^*_{\mu 2} = -0.224 - 0.11i \qquad  
U_{\tau 1} U^*_{\tau 2} = -0.224 + 0.11i  \label{fir}
\end{equation}
where two sides are equal in length and the internal angles of the triangle are
$26.1^{\circ}$ (for two of the angles) and $127.8^{\circ}$.
For the second Majorana triangle we have
\begin{equation}
U_{e1} U^*_{e3} = 0.175i \qquad
U_{\mu 1} U^*_{\mu 3} = - 0.2815 - 0.0875 i \qquad
U_{\tau 1} U^*_{\tau 3} = 0.2815 - 0.0875 i   \label{sec}
\end{equation}
once again two of the sides are equal in length. In this case two 
of the internal
angles are equal to $72.7^{\circ}$ and the other one to $34.6 ^{\circ}$. 
Finally for the third Majorana triangle we have
\begin{equation}
U_{e2} U^*_{e3} = 0.1238i \ \ \ 
U_{\mu 2} U^*_{\mu 3} = 0.3980 - 0.0619i \ \ \   
U_{\tau 2} U^*_{\tau 3} = -0.3980 + 0.0619i
\end{equation}
Two sides have equal length, leading to two internal angles
of $81.2^{\circ}$ and another angle of $17.6 ^{\circ}$. Of the 
three triangles thus obtained, this is the one  
with a smallest internal angle. 
Note that all three triangles are isosceles, which
results from the fact 
there is equality of moduli between rows two and three
of the mixing matrix. This due to the particular values
of $\theta_{23}$ and $\delta$.
Perturbations around the
HPS values for $\theta_{12}$ and $\theta_{23}$ in the range
still allowed by experiment would not alter significantly
the shape of these triangles.

An alternative generalization of the tri-bimaximal form was 
considered in reference \cite{Bjorken:2005rm} where the exact 
trimaximal mixing of the second column is maintained
and unitarity is imposed by construction, with $U_{e3}$ now different
from zero and possibly complex. In this construction small
deviations from the HPS values of  $\theta_{12}$ and $\theta_{23}$
occur and $\mu$ - $\tau$ reflection symmetry
\cite{Lam:2001fb}, \cite{Harrison:2002et} is broken for Re($U_{e3}$)  
different from zero.  The Majorana-type triangles thus obtained, involving
orthogonality relations with the second column become specially
simple. The third Majorana triangle has the interesting feature 
of one of the sides being simply proportional  $U_{e3}$. 
In this approach $\theta_{13}$ and $\theta_{12}$ are related 
by the constraint of trimaximal mixing in the second column.
For the maximal $\theta_{13}$ still allowed experimentally
($\sin \theta_{13} = 0.22$) together with $|U_{e2}|$ fixed
as $ 1/ \sqrt{3}$ which is a consequence of imposing  trimaximal mixing 
in the second column, and with unitarity
we are lead to:
\begin{equation}
\sin^2 (2 \theta_{12}) = 0.91
\end{equation}
a value for $\theta_{12}$ which is already disfavoured, as can be
seen from Eq.~(\ref{sim}). \\

So far, in this section  we assumed unitarity of the PMNS matrix, 
together with the presence of Dirac-type CP violation, 
which in turn allowed for its reconstruction from
six Majorana phases. Yet, it should be noted that 
deviations from unitarity naturally arise in a variety of 
extensions of the SM, involving the lepton and or quark sectors.
Actually, in the context of standard seesaw the $3 \times 3$ PMNS
matrix is not exactly unitary. However in this 
framework deviations from unitarity cannot be detected experimentally
due to the extreme degree of their suppression.
On the other hand there are extensions of the SM where 
experimentally detectable deviations from unitarity may arise. 
Examples include models with 
vector-like quarks \cite{delAguila:1985mk} --- \cite{Botella:2008qm}
as  well as models with heavy Majorana neutrinos
with masses of order 1 TeV or lower
\cite{Bray:2007ru}, \cite{del Aguila:2007em}.  
Majorana neutrino singlets
with no new gauge interactions might be produced 
within the reach of the LHC, up to masses of 
order 200 Gev \cite{del Aguila:2007em}. 
The possibility of having extensions of the SM with natural violations
of $3 \times 3$ unitarity raises the question of how to test experimentally 
the validity of the unitarity hypothesis.
A set of exact relations connecting measurable quantities were derived 
for the quark sector \cite{Botella:2002fr}
providing tests of unitarity of the $V_{CKM}$ matrix. Similar relations 
can be derived in the leptonic sector. Examples of such relations,
derived from the Majorana-type triangles, are:
\begin{eqnarray}
\frac{|U_{e1} U^*_{e2}|}{\sin(\beta_2 - \beta_3)} 
=  \frac{|U_{\mu 1} U^*_{\mu 2}|} {\sin(\beta_1 - \beta_3)}
=\frac{| U_{\tau 1} U^*_{\tau 2}|}{\sin(\beta_1 - \beta_2)}  
\label{um}
\end{eqnarray}
\begin{eqnarray}
\frac{|U_{e1} U^*_{e3}|}{\sin(\gamma_2 - \gamma_3)} 
= \frac{| U_{\mu 1} U^*_{\mu 3}|}{\sin(\gamma_1 - \gamma_3)}
=\frac{| U_{\tau 1} U^*_{\tau 3}|}{\sin(\gamma_1 - \gamma_2)}
\label{do} 
\end{eqnarray}
\begin{eqnarray}
\frac{|U_{e2} U^*_{e3}|}{\sin(-\beta_2 + \beta_3 + \gamma_2 - \gamma_3)}  
= \frac{| U_{\mu 2} U^*_{\mu 3}|}{\sin(-\beta_1 + \beta_3 + \gamma_1 - 
\gamma_3)}
= \frac{| U_{\tau 2} U^*_{\tau 3}|}{\sin(-\beta_1 + \beta_2 + \gamma_1 - 
\gamma_2)}
\label{tr}
\end{eqnarray}
with analogous relations for the Dirac-type triangles. 
Although relations (\ref{um}),  (\ref{do}),  (\ref{tr}) are exact 
predictions of the PMNS framework, relating physically measurable quantities,
their  experimental test is a great challenge which
would require the experimental discovery of leptonic 
CP violation of  Dirac-type \cite{Farzan:2006vj},
\cite{FernandezMartinez:2007ms}

Whenever the length of the largest side of the triangle is smaller 
than the sum of the lenghts of the other two, several possibilities
arise. Either there is Dirac-type CP violation or violation of 
unitarity of the PMNS matrix or both. In reference \cite{Farzan:2002ct}
a set of measurements is suggested which will, in principle, allow 
to measure all sides of the e-$\mu$ Dirac unitarity triangle. \\

We have previously emphasized that the orientation of Majorana
triangles has physical meaning since they are related to
the size of certain Majorana-type phases. This raises the  question of 
which observables would in principle be sensitive to these
orientations. It is well known that neutrino oscillations
are only sensitive  to Dirac-type CP violation
and thus its experimental discovery only provides information
about differences of Majorana phases, like  $\beta_1 -\beta_2$
or $\gamma_1 -\gamma_2$ but not on the individual values of
$\beta_i$, $\gamma_i$. As a result no knowledge about the orientation 
of Majorana triangles can be obtained from the detection of
Dirac-type CP violation. 

In the next sections we discuss the question of how neutrinoless 
double beta decay as well as leptogenesis \cite{Fukugita:1986hr}
when flavour effects matter \cite{Barbieri:1999ma}, \cite{Endoh:2003mz},
\cite{Fujihara:2005pv},
\cite{Abada:2006fw}, \cite{Nardi:2006fx}, \cite{Abada:2006ea} 
are sensitive to the Majorana-type phases.

\section{Majorana phases and the elements of the neutrino effective 
mass matrix}
In the leptonic low energy limit and in the weak  
basis where the mass matrix of the charged leptons is real and diagonal, 
the effective neutrino mass matrix $m_{eff}$
is complex, symmetric, with nine independent
parameters. Although it may in principle be fully reconstructed from experiment, 
it has been pointed out that it is not possible, in practice, to fully
reconstruct $m_{eff}$ without ambiguities, from a set of feasible 
experiments. This has motivated several authors to 
introduce some input from theory in order to allow for this
reconstruction \cite{Frampton:2002yf},
\cite{Branco:2002ie}. \\

In the seasaw framework the effective Majorana mass matrix is given by 
\begin{equation}
m_{eff}= - m_D \frac{1}{M_R}{m_D}^T 
\label{eq4}
\end{equation}
where $m_D$ is the Dirac-type mass matrix and $M_R$ is the Majorana mass matrix
for the righthanded neutrino singlets. With this notation the connection
among light neutrino masses and the elements of the PMNS matrix, starting from the 
weak basis specified above, is established through the relation:
\begin{equation}
U^{\dagger}m_{eff}U^* = d = {\rm diag} \ (m_1, m_2, m_3)
\label{mdu}
\end{equation}
From this equation it is clear that each entry of $m_{eff}$, to be denoted
in what follows by $m_{ij}$,
can be fully expressed in terms of observable quantities
-- neutrino masses, mixing angles and phases.  The 
absolute value of the element (11) of
$m_{eff}$ is specially interesting experimentally since, in
the absence of additional lepton number violating interactions other than  
those generated by the charged currents involving Majorana neutrinos, 
it can be measured in neutrinoless double beta 
decay experiments  \cite{KlapdorKleingrothaus:2000sn}, \cite{Aalseth:2002rf},
\cite{KlapdorKleingrothaus:2004wj}.

From Eq.~(\ref{mdu}) we obtain:
\begin{eqnarray}
|m_{11}|^2 = m^2_1 |U_{e1}|^4 +  m^2_2 |U_{e2}|^4+
 m^2_3 |U^2_{e3}|^4 + 2 m_1 m_2 |U_{e1}|^2 |U_{e2}|^2 \cos(2 \beta_1)+ 
\nonumber \\ 
+ 2 m_1 m_3 |U_{e1}|^2 |U_{e3}|^2 \cos(2 \gamma_1) +
+ 2 m_2 m_3 |U_{e2}|^2 |U_{e3}|^2 \cos[2 (\beta_1-\gamma_1)]
\label{111}
\end{eqnarray}
the angle $(\gamma_1 - \beta_1)$ is the argument of $U^*_{e1} U_{e2} U_{e1}
U^*_{e3}$, which is not a rephasing invariant Dirac-type quartet.
The corresponding product in the 
quark sector, in terms of elements of $V_{CKM}$, would not be a 
rephasing invariant. It is the Majorana character of the neutrinos  
that gives physical meaning to the phase of this fourfold product.
If we were to rewrite Eq.~(\ref{111}) using the parameterisation of the PMNS
matrix given by Eq.~(\ref{pdg}) the Dirac phase $\delta$ would appear 
explicitly. On the other hand, it is always possible to eliminate 
the explicit dependence on $\delta$
from $|m_{11}|$  by redefining the factorizable phase $\beta$ in such a 
way that the phase $\delta$ 
only appears on the second and third rows of the PMNS matrix.
This may seem paradoxical, but it has a simple explanation.
There is Dirac-type
CP violation only when the PMNS matrix contains non-factorizable
Majorana-type phases.
The measurement of $|m_{11}|$ is only sensitive to one row 
of the PMNS matrix. When one single row of the PMNS matrix
is considered it is always possible to factor out all physical phases on
the righthand side of the matrix. It is necessary to combine information
from other rows in order to extract 
information on the possible presence of non-factorizable phases. 
Provided we know the masses 
of each of the light neutrinos, once we measure the modulus of  $m_{11}$
we can infer whether or not there are relative phases among
each term and therefore whether or not there is Majorana-type CP violation.

Unfortunately, there are no known feasible experiments that would allow 
us to measure directly the modulus of other entries of $m_{eff}$.
For the off-diagonal entries we have:
\begin{eqnarray}
|m_{ij}|^2 =  m^2_1 |U_{i1}|^2  |U_{j1}|^2 +  
m^2_2  |U_{i2}|^2  |U_{j2}|^2 + m^2_3  |U_{i3}|^2  |U_{j3}|^2+  \nonumber \\ 
+ 2 m_1 m_2 |U_{i1}| |U_{i2}| |U_{j1}| |U_{j2}|\cos( \beta_i+ \beta_j)+ 
\nonumber \\ 
+ 2 m_1 m_3  |U_{i1}| |U_{i3}| |U_{j1}| |U_{j3}|\cos(\gamma_i+ \gamma_j)+
\nonumber \\ 
+ 2 m_2 m_3  |U_{i2}| |U_{i3}| |U_{j2}| |U_{j3}|
\cos(\gamma_i+ \gamma_j -\beta_i- \beta_j)
\end{eqnarray}
This expression combines information involving two rows of the
PMNS matrix where again the Majorana-type phases appear in combinations
that are not Dirac-type and that would not be rephasing invariant
for Dirac neutrinos. The measurement of one of the $|m_{ij}|$,
together with the knowledge of the three neutrino masses, would only
give information on the sum of two  $\beta_j$ and the sum of two  $\gamma_j$
without  allowing to determine whether or not the Majorana phases
are factorizable.

\section{Majorana phases and CP-odd Weak Basis invariants}

We have seen that
leptonic CP violation at low energies requires the presence of complex
Majorana-type bilinears, which are  defined in terms of entries of the
PMNS matrix. The information on whether or not a
Lagrangian violates CP is also encoded in the fermionic
mass matrices  written in a weak basis. Unlike the 
physical basis, weak bases are not unique and, as a result,
there is an infinite number of sets of fermion mass matrices
corresponding to the same physics. It is often practical 
to analyse the CP properties of the Lagrangian in terms of  CP-odd
weak basis invariants. Different WB invariants
are sensitive to different CP violating phases in different
physical scenarios. The general strategy to build such WB invariants
was outlined for the first time in 
Ref.~\cite{Bernabeu:1986fc} and in Ref.~\cite{Branco:2004hu}
several relevant examples are given together with
additional references. \\

The strength of  Dirac-type CP violation can be obtained from the
following low energy WB invariant:
\begin{equation}
Tr[h_{eff}, h_l]^3= - 6i \Delta_{21} \Delta_{32} \Delta_{31}
{\rm Im} \{ (h_{eff})_{12}(h_{eff})_{23}(h_{eff})_{31} \} \label{trc}
\end{equation}
where $h_{eff}=m_{eff}{m_{eff}}^{\dagger} $, 
$ h_l = m_l m^\dagger_l $, and
$\Delta_{21}=({m_{\mu}}^2-{m_e}^2)$ with analogous expressions for
$\Delta_{31}$, $\Delta_{32}$. The righthand side of this equation
is the computation of this invariant in the special WB where the 
charged lepton masses are real and diagonal. An analogous invariant  
is relevant for the quark sector \cite{Bernabeu:1986fc}.
This WB invariant can  be fully expressed in terms of physical 
observables since
\begin{equation}
{\rm Im} \{ (h_{eff})_{12}(h_{eff})_{23}(h_{eff})_{31} \} =
- \Delta m_{21}^2 \Delta m_{31}^2  \Delta m_{32}^2 {\rm Im} Q
\label{fjcp} 
\end{equation}
where $\mbox{Im}Q$ is the imaginary part 
of a rephasing invariant quartet of the leptonic mixing matrix $U$
and signals the presence of Dirac-type CP violation.

It is also possible to construct 
WB invariants that are sensitive to Majorana-type phases. 

It has been shown \cite{Branco:1986gr} that
the condition
\begin{equation}
{\rm Im \ tr } \; F = 0 \label{trq}
\end{equation}
with $F = h_l  m_{eff}   m^*_{eff}  m_{eff} h^*_l  m^*_{eff}$
is a necessary condition for CP invariance in the
leptonic sector, for an arbitrary number of light Majorana neutrinos.
This CP odd invariant is sensitive to Majorana-type phases
and it may not vanish even in the case where there is 
no Dirac-type CP violation.In order to see that 
this is the case, it is useful to compute it 
in terms of lepton masses, mixing angle and 
CP violating phase
in the  simple case of two generations, where there is no
Dirac-type CP violation but Majorana-type CP violation occurs. One
obtains:
\begin{equation}
{\rm Im \ tr } \; F = \frac{1}{4} m_1m_2\ (m^2_2 -m^2_1)
(m^2_{\mu}-m^2_{e})^2 \sin^2 \ 2 \theta \sin^2 \ 2 \gamma 
\end{equation}
where the $2 \times 2$ 
leptonic mixing matrix is parameterised as:
\begin{equation}   
K = \left[
\begin{array}{cc}
\cos \theta  & - \sin \theta e^{i \gamma} \\
 \sin \theta e^{- i \gamma}  &  \cos \theta  
\end{array}
\right]
\label{old}
\end{equation} 
It is the Majorana character of the neutrinos that prevents 
the phase $\gamma$ in Eq.~(\ref{old}) to be rotated away.
The phase $(- \gamma)$ is in fact the argument of the Majorana
bilinears $(K_{11} K^*_{12})$ and $(K_{21} K^*_{22})$, modulo $\pi$.

Another peculiar aspect of Majorana neutrinos is the fact that 
for three Majorana neutrinos there is CP violation even in the limit
of exact degeneracy of neutrino masses. In this limit,
a necessary and sufficient condition 
\cite{Branco:1998bw} for CP invariance is:
\begin{equation}
G\equiv \ {\rm {Tr}}\left[ \ m^*_{eff}\cdot h_l\cdot m_{eff}\ ,\
h^*_l\right] ^3\ =\ 0.
\end{equation}
Therefore, this WB invariant condition must be sensitive to 
Majorana-type CP violation even in the absence of Dirac-type 
CP violation, both in the case of degenerate and nondegenerate 
neutrino masses. 
By analogy to Eq.~(\ref{trc}) we may write:
\begin{eqnarray}
& G = - 6i \Delta_{21} \Delta_{32} \Delta_{31} \times \nonumber \\
& \times {\rm Im} \{ {(m^*_{eff}\cdot h_l\cdot m_{eff})}_{12}
{(m^*_{eff}\cdot h_l\cdot m_{eff})}_{23}
{(m^*_{eff}\cdot h_l\cdot m_{eff})}_{31} \} \label{xyz}
\end{eqnarray}
It can be checked that $G$ is indeed sensitive to Majorana bilinears,
by writing each factor of the form 
${(m^*_{eff}\cdot h_l\cdot m_{eff})}_{ij}$, $i \neq j$,
explicitly in terms of masses and mixing, with the help 
of Eq.~(\ref{mdu}).
It is the presence of the matrix $h_l$ between $m^*_{eff}$ 
and $m_{eff}$ that makes this CP odd invariant 
fundamentally different from
the one in Eq.~(\ref{trc}). The terms in $m_i m_j$ with  $i \neq j$
which are generated once we expand the above factors, always appear 
multiplied by Majorana bilinears and also by the square 
of a charged lepton
mass. These three different factors prevent the possibility of 
simplification among these terms which otherwise would add to zero
due to unitarity.
If the charged leptons were degenerate in mass 
only the terms in $m^2_j$ of the expansion,
would survive. These terms do not depend on Majorana bilinears.

\section{Majorana phases and Leptogenesis}
CP violation in the leptonic sector may play a 
fundamental r\^ ole in the generation, via leptogenesis,
of the observed baryon number asymmetry of the 
universe (BAU)\cite{Bennett:2003bz}:
\begin{equation}
\frac{n_{B}}{n_{\gamma}}= (6.1 ^{+0.3}_{-0.2}) \times 10^{-10}.
\end{equation}
In this framework a CP asymmetry is generated through
out-of-equilibrium L-violating decays of heavy Majorana
neutrinos \cite{Fukugita:1986hr} leading to a lepton asymmetry
which, in the presence of  $(B+L)$-violating but 
$(B-L)$-conserving sphaleron processes \cite{Kuzmin:1985mm},
produces a baryon asymmetry.

In the single flavour approach, with three singlet heavy neutrinos
$N_i$, thermal leptogenesis is insensitive to the CP violating
phases appearing in the PMNS matrix. In this case there is
complete decoupling among the phases responsible for 
CP violation at low energies and those
responsible for leptogenesis \cite{Branco:2001pq},\cite{Rebelo:2002wj}.

From Eq.~(\ref{mdu}) and the definition of $m_{eff}$ one can write $m_D$
in the Casas and Ibarra parameterisation \cite{Casas:2001sr} as:
\begin{equation}
m_D = i U {\sqrt d} R {\sqrt D}
\label{udr}
\end{equation}
The matrix $R$ is a general complex orthogonal matrix, and $d$ and $D$
are diagonal matrices for the light and the heavy
neutrino masses, respectively. Clearly low 
energy physics cannot provide any information on $R$ since
this matrix cancels out in $m_{eff}$. 
The lepton number asymmetry resulting from
the decay  of heavy Majorana neutrinos, $\varepsilon _{N_{j}}$,
was computed, in the single flavour approach,
by several authors \cite{Covi:1996wh}, \cite{Flanz:1994yx}, 
\cite{Plumacher:1996kc}. The result is proportional to 
$\sum_{k \ne j}{\rm Im} (m_D^\dagger m_D)_{jk} (m_D^\dagger m_D)_{jk}$
with an additional factor depending on the ratio of the masses 
of the two heavy neutrinos $k$ and $j$,   $x_k=\frac{{M_k}^2}{{M_j}^2}$.
The matrix $U$ cancels out in the combination  $m_D^\dagger m_D$
and in this case leptogenesis only depends on CP violation present 
in $R$. This is a consequence of  having summed up into all 
charged lepton indices  $l_i^\pm$ ($i$ = e, $\mu$ , $\tau$) resulting 
from the decay of the heavy neutrino.

Flavour effects matter when washout processes are sensitive to 
the different leptonic flavours produced in the decay of heavy 
Majorana neutrinos \cite{Davidson:2008bu}.
In this particular case the single flavour approach
ceases to be valid and the separate asymmetry produced in each decay
has to be considered.

The separate lepton $i$ family asymmetry  $\varepsilon^i _{N_j}$
generated from the decay 
of the $j$th heavy Majorana neutrino
is given by \cite{Endoh:2003mz}:
\begin{eqnarray}
\varepsilon^i _{N_j} = \frac{g^2}{M^2_W} \frac{1}{16\pi}
\sum_{k \ne j} \left[ I(x_{k})\frac{\mbox{Im}\left(
 (m_D^\dagger m_D)_{jk}(m_D^*)_{ij} (m_D)_{ik}\right)}
{\left| (m_D)_{ij} \right|^2 } \right.  \nonumber \\
+ \left. \frac{1}{1 - x_{k}} \frac{ \mbox{Im} \left(
 (m_D^\dagger m_D)_{kj}(m_D^*)_{ij} (m_D)_{ik}\right) }
{\left|(m_D)_{ij}\right|^2 } \right] 
\label{fla}
\end{eqnarray}
with 
\begin{equation}
I(x_{k}) = \sqrt{x_k}\left[ 1+ \frac{1}{1-x_k} + 
(1+x_k) \ln \frac{x_k}{1+x_k} \right]
\end{equation}
Clearly, when one works with separate flavours the matrix $U$ 
does not cancel out and one is lead to the interesting possibility 
of having viable leptogenesis even in the case of $R$ being a 
real matrix \cite{Branco:2006hz},
\cite{Pascoli:2006ie}, \cite{Branco:2006ce}, \cite{Uhlig:2006xf}.
If we were to sum over all charged leptons, the first 
term in  Eq.~(\ref{fla}) would lead to the expression
obtained for the total lepton number asymmetry in the case of
unflavoured leptogenesis, whilst the second term would become real.

Assuming $R$ to be real, from Eq.~(\ref{udr}) we obtain:
\begin{eqnarray}
& \mbox{Im}\left( (m_D^\dagger m_D)_{jk}(m_D^*)_{ij} (m_D)_{ik}\right) 
= \nonumber \\
& = (m_D^\dagger m_D)_{jk} {\sqrt d_l} R_{lj} {\sqrt D_j}
{\sqrt d_s} R_{sk} {\sqrt D_k}|U_{il}| |U_{is}| 
\sin \left( \arg (U^*_{il}U_{is}) \right)
\end{eqnarray}
The only indices that are summed up are 
$l$ and $s$ and each term in this sum is proportional to
the sine of a $\beta_i$, or a $\gamma_i$, or a $(\beta_i - \gamma_i)$,
which are  pure Majorana-type phases.  The second term, 
Im$\left( (m_D^\dagger m_D)_{kj}(m_D^*)_{ij} (m_D)_{ik}\right) $,
only differs from this one by the structure of indices
of $(m_D^\dagger m_D)$.
Flavoured leptogenesis is 
sensitive to each one of the different Majorana-type phases
alone and, in the general case of complex $R$, it will depend on
the additional phases present in this matrix. 

\section{Conclusions}
We have emphasized that in the case of Majorana neutrinos, 
the arguments of rephasing invariant bilinears, designated 
Majorana-type phases, are the fundamental quantities in 
the study of CP violation in the leptonic sector. If
one further assumes $3 \times 3$ unitarity of the PMNS
matrix we have shown that in general the full PMNS matrix can be 
derived using as input six independent Majorana-type phases.
The presence of non factorizable 
Majorana-type  phases in the PMNS matrix signals the presence of Dirac-type
CP violation which might be observable in future
neutrino oscillation experiments.
As a result, Dirac-type CP violation requires the existence of 
Majorana-type CP violation. Obviously the converse is not true.
We have  shown how to relate the strength of Dirac-type
CP violation to these Majorana-type phases by writing the area of 
the unitarity triangles in terms of these phases.
We have also studied how these Majorana-type phases
appear in the elements of the neutrino mass matrix, as
well as in flavoured leptogenesis.

   Observables that should be sensitive to the Majorana-type
phases, even in the absence of Dirac-type CP violation, include
neutrinoless double beta decay and possibly leptogenesis. 
Neutrino-antineutrino oscillation processes can also in 
principle be used to measure CP-violating Majorana phases
\cite{deGouvea:2002gf}. Other manifestely CP violating
physical processes are leptonic electric dipole moments
\cite{Mohapatra:1998rq}. An extensive review of issues related
to flavour phenomena and CP violation in the leptonic sector
and the potential for their discovery in the LHC and
possible future experiments is provided in Ref.~\cite{Raidal:2008jk}

   It is clear that the application of our results to perform 
practical tests of the PMNS paradigm is severely
restricted by the scarcity of data on leptonic mixing and
CP violation, leading to the dreadful situation that 
the neutrino mass matrix cannot be fully reconstructed 
from a set of presently conceived feasible experiments.
One possible hope is having a significant development
in our understanding of flavour, in particular of 
leptonic flavour. If a theory of flavour implies, for example,
direct constraints on the Majorana-type phases, then the 
relations we have derived, connecting these phases 
to other leptonic observables, would be of paramount importance.

\section*{Acknowledgements}
This work was partially supported by Funda\c c\~ ao para a 
Ci\^ encia e a  Tecnologia (FCT, Portugal) through the projects
PDCT/FP/63914/2005, PDCT/FP/63912/2005, \\
POCI/81919/2007,
and CFTP-FCT UNIT 777  which are partially funded through POCTI 
(FEDER) and by the Marie Curie RTNs MRTN-CT-2006-035505 and
MRTN-CT-503369. G.C.B. would
like to thank Francisco Botella for the kind hospitality at
Universitat de Val\` encia-CSIC during his sabbatical leave. 
MNR would like to thank 
Francisco Botella for the warm welcome during the short visit
to Universitat de Val\` encia supported by Ac\c c\~ oes de Mobilidade
Portugal/Espanha 2008.
The authors are grateful for the warm 
hospitality of the CERN Physics Department (PH) Theory Division (TH) 
where part of this work was done.

\end{document}